\documentclass{article}
\usepackage{amsmath}
\usepackage{amsfonts}
\newcommand{\bosy}[1]{\boldsymbol{#1}}
\DeclareMathOperator{\arc}{arc}
\title{Looking for a time independent Hamiltonian of a dynamical system}
\author{Micha{\l} Dobrski
\footnote{mdobrski@im$\emptyset$.p.lodz.pl} 
\\
\small
\emph{Center of Mathematics and Physics,}
\\
\small
\emph{Technical University of {\L}\'od\'z,}
\\
\small
\emph{Al.~Politechniki~11, 90-924~{\L}\'od\'z, Poland}
}
\date{}
\begin{document}
\maketitle
\begin{abstract}
In this paper we introduce a method for finding a time independent Hamiltonian of a given dynamical system by canonoid transformation of canonical momenta. We find a condition that the system should satisfy to have an equivalent time independent formulation. We study the example of damped oscillator and give the new time independent Hamiltonian for it, which has the property of tending to the standard Hamiltonian of the harmonic oscillator as damping goes to zero.
\end{abstract}
\section{Intoduction}
It is known that the Hamiltonian description of the system is nonunique. Generally, one can change phase-space coordinates ($\bosy{q}$, $\bosy{p}$) and in some cases it is possible to find a Hamiltonian for these new coordinates. The well known class of such transformations are \emph{canonical transformations} which preserve Hamiltonian structure for \emph{all} possible Hamiltonians (\cite{gantmacher}, \cite{arnold}). But there is also much wider class of transformations, called \emph{canonoid} which preserve \emph{at least one} Hamiltonian structure. Theory of time canonoid transformations is frequently investigated in a context of inverse problem of the calculus of variations (\cite{santilli2}, \cite{morandi} and references there). General theory of time independent canonoid transformations can be found in \cite{santilli2}. 

The main goal of this paper is to develop a method for finding a time independent Hamiltonian for a given system. We restrict ourselves to transformations which do not change $\bosy{q}$, because we could have some physical (e.g. quantum) reasons for preserving configurational coordinates. (Without this restriction the problem could be trivial, as the transformation to coordinates of integrals of motion would yield the Hamiltonian constantly equal to zero). This kind of canonoid transformation sometimes is called \emph{fouling transformation} (\cite{gelmansaletan}).

The paper is organized as follows. First, (section 2) as an introduction, we present theory of canonoid transformations preserving $\bosy{q}$. Then (section 3) we find additional relations which the transformation must satisfy to provide the transition to phase space of the time independent Hamiltonian. We show that such a transition is not possible for all Hamiltonians, and we find equation which is necessary condition for the system to have a time independent Hamiltonian. Finally (section 4), we test all this methods on the damped oscillator. We find the time independent Hamiltonian for this system, which tends to the classical Hamiltonian of the harmonic oscillator as damping goes to zero.

\section{Canonoid transformations preserving $\bosy{q}$}
Consider two dynamical systems. One described by coordinates $(\bosy{q}$, $\bosy{p},t)$ and the Hamiltonian $H_0(\bosy{q},\bosy{p},t)$, and the other expressed by coordinates ($\bosy{q}$, $\bosy{\pi},t)$ and the Hamiltonian $H(\bosy{q},\bosy{\pi},t)$. Let us also have the mapping $\bosy{\tilde{p}}(\bosy{q},\bosy{p},t)$ between momenta $\bosy{p}$ and $\bosy{\pi}$, such that $(\bosy{q},\bosy{p},t) \mapsto (\bosy{q},\bosy{\tilde{p}}(\bosy{q},\bosy{p},t),t)$ is diffeomorphism. We will call $H_0$ and $H$ equivalent if each solution of Hamilton equations for $H_0$ is also (by means of $\bosy{\tilde{p}}$) solution of Hamilton equations for $H$ and vice versa. One can easily check that if both Hamiltonians lead to unique solutions (in terms of initial conditions) then it is sufficient only to check the transition from $H_0$ to $H$. Condition of equivalence for the inverse transformation is then automatically fullfiled (at least locally).

The Hamilton equations for $H_0$ read
\begin{equation}
\label{rowhamstare}
\frac{\textrm{d} q_i}{\textrm{d} t}(t)=\left.\frac{\partial H_0}{\partial p_i}\right|_{\substack{\bosy{q}=\bosy{q}(t)\\\bosy{p}=\bosy{p}(t)}}, \qquad 
\frac{\textrm{d} p_i}{\textrm{d} t}(t)=-\left.\frac{\partial H_0}{\partial q_i}\right|_{\substack{\bosy{q}=\bosy{q}(t)\\\bosy{p}=\bosy{p}(t)}}.
\end{equation}
Now $H$ is equivalent to $H_0$ if for every solution $\bosy{q}(t)$, $\bosy{p}(t)$ of these equations, $\bosy{q}(t)$ and $\bosy{\pi}(t)=\bosy{\tilde{p}}(\bosy{q}(t),\bosy{p}(t),t)$ are solutions of Hamilton equations for $H$
\begin{equation}
\label{rowhamnowe}
\frac{\textrm{d} q_i}{\textrm{d} t}(t)=\left.\frac{\partial H}{\partial \pi_i}\right|_{\substack{\bosy{q}=\bosy{q}(t)\\\bosy{\pi}=\bosy{\pi}(t)}}, \qquad 
\frac{\textrm{d} \pi_i}{\textrm{d} t}(t)=-\left.\frac{\partial H}{\partial q_i}\right|_{\substack{\bosy{q}=\bosy{q}(t)\\\bosy{\pi}=\bosy{\pi}(t)}}.
\end{equation}
Comparing these two sets of equations one gets the following relations
\begin{equation*}
\left.\frac{\partial H_0}{\partial p_i}\right|_{\substack{\bosy{q}=\bosy{q}(t)\\ \bosy{p}=\bosy{p}(t)}}=
\left.\frac{\partial H}{\partial \pi_i}\right|_{\begin{subarray}{l}\bosy{q}=\bosy{q}(t)\\ \bosy{\pi}=\bosy{\tilde{p}}(\bosy{q}(t),\bosy{p}(t),t)\end{subarray}}
\end{equation*}
and
\begin{equation*}
\left.
\{\tilde{p}_j,H_0 \}
\right|_{\substack{\bosy{q}=\bosy{q}(t)\\\bosy{p}=\bosy{p}(t)}} +
\left.\frac{\partial \tilde{p}_j}{\partial t}\right|_{\substack{\bosy{q}=\bosy{q}(t)\\\bosy{p}=\bosy{p}(t)}}
=
-\left.\frac{\partial H}{\partial q_j}\right|_{\begin{subarray}{l}\bosy{q}=\bosy{q}(t)\\\bosy{\pi}=\bosy{\tilde{p}}(\bosy{q}(t),\bosy{p}(t),t)\end{subarray}}.
\end{equation*}
Then, in terms of the coordinates, we obtain
\begin{equation}
\label{dosycwazne}
\left.\frac{\partial H_0}{\partial p_i}\right|_{\begin{subarray}{1}\bosy{q}=\bosy{q}\\ \bosy{p}=\bosy{p}\end{subarray}}=
\left.\frac{\partial H}{\partial \pi_i}\right|_{\begin{subarray}{1}\bosy{q}=
\bosy{q}\\\bosy{\pi}=\bosy{\tilde{p}}(\bosy{q},\bosy{p},t)\end{subarray}}
\end{equation}
and
\begin{equation}
\label{dosycwaznedrugie}
-\left
\{\tilde{p}_j,H_0 \}
\right|_{\substack{\bosy{q}=\bosy{q}\\\bosy{p}=\bosy{p}}}
-\left.
\frac{\partial \tilde{p}_j}{\partial t}
\right|_{\substack{\bosy{q}=\bosy{q}\\\bosy{p}=\bosy{p}}}
=
\left.\frac{\partial H}{\partial q_j}\right|_{\begin{subarray}{1}\bosy{q}=
\bosy{q}\\\bosy{\pi}=\bosy{\tilde{p}}(\bosy{q},\bosy{p},t)\end{subarray}}.
\end{equation}
These are our fundamental relations and we can treat them as the equations for the new Hamiltonian $H$ and for the mapping $\bosy{\tilde{p}}$. We can easily transform them to the following form
\begin{equation}
\label{wazne1}
\left.\frac{\partial H_0}{\partial p_i}\right|_{\begin{subarray}{1}\bosy{q}=\bosy{q}\\ \bosy{p}=\bosy{p}\end{subarray}} \frac{\partial \tilde{p}_i}{\partial p_j} =
\frac{\partial}{\partial p_j} \left(
\left.H\right|_{\begin{subarray}{1}\bosy{q}=
\bosy{q}\\\bosy{\pi}=\bosy{\tilde{p}}(\bosy{q},\bosy{p},t)\end{subarray}} \right)
\end{equation}
and similarly
\begin{equation}
\label{wazne2}
-\left.
\{\tilde{p}_j,H_0 \}
\right|_{\substack{\bosy{q}=\bosy{q}\\\bosy{p}=\bosy{p}}}
-\left.
\frac{\partial \tilde{p}_j}{\partial t}
\right|_{\substack{\bosy{q}=\bosy{q}\\\bosy{p}=\bosy{p}}}
+
\left.\frac{\partial H_0}{\partial p_i}\right|_{\begin{subarray}{1}\bosy{q}=\bosy{q}\\ \bosy{p}=\bosy{p}\end{subarray}} \frac{\partial \tilde{p}_i}{\partial q_j}
=
\frac{\partial}{\partial q_j} \left(
\left.H\right|_{\begin{subarray}{1}\bosy{q}=
\bosy{q}\\\bosy{\pi}=\bosy{\tilde{p}}(\bosy{q},\bosy{p},t)\end{subarray}} \right).
\end{equation}
Above equations can be viewed as definitions for the derivatives of $\left.H\right|_{\begin{subarray}{1}\bosy{q}=
\bosy{q}\\\bosy{\pi}=\bosy{\tilde{p}}(\bosy{q},\bosy{p},t)\end{subarray}}$. But such definition could make sense if the integrability conditions
\begin{subequations}
\begin{gather}
\label{warcalk1}
\frac{\partial^2}{\partial q_j \partial q_k} \left(
\left.H\right|_{\begin{subarray}{1}\bosy{q}=
\bosy{q}\\\bosy{\pi}=\bosy{\tilde{p}}(\bosy{q},\bosy{p},t)\end{subarray}} \right)
-
\frac{\partial^2}{\partial q_k \partial q_j} \left(
\left.H\right|_{\begin{subarray}{1}\bosy{q}=
\bosy{q}\\\bosy{\pi}=\bosy{\tilde{p}}(\bosy{q},\bosy{p},t)\end{subarray}} \right)
=0,\\
\frac{\partial^2}{\partial p_j \partial p_k} \left(
\left.H\right|_{\begin{subarray}{1}\bosy{q}=
\bosy{q}\\\bosy{\pi}=\bosy{\tilde{p}}(\bosy{q},\bosy{p},t)\end{subarray}} \right)
-
\frac{\partial^2}{\partial p_k \partial p_j} \left(
\left.H\right|_{\begin{subarray}{1}\bosy{q}=
\bosy{q}\\\bosy{\pi}=\bosy{\tilde{p}}(\bosy{q},\bosy{p},t)\end{subarray}} \right)
=0,\\
\frac{\partial^2}{\partial p_j \partial q_k} \left(
\left.H\right|_{\begin{subarray}{1}\bosy{q}=
\bosy{q}\\\bosy{\pi}=\bosy{\tilde{p}}(\bosy{q},\bosy{p},t)\end{subarray}} \right)
-
\frac{\partial^2}{\partial q_k \partial p_j} \left(
\left.H\right|_{\begin{subarray}{1}\bosy{q}=
\bosy{q}\\\bosy{\pi}=\bosy{\tilde{p}}(\bosy{q},\bosy{p},t)\end{subarray}} \right)
=0
\label{warcalk3}
\end{gather}
\end{subequations}
are fulfilled. This yields the following equations
\begin{subequations}
\begin{align}
\label{wynik1}
[q_j,q_k]_{(\tilde{p}_m, \dot{q}_m)} &=
\frac{\partial}{\partial q_k}\left(\{\tilde{p}_j,H_0 \}
+\frac{\partial \tilde{p}_j}{\partial t}
\right)
-
\frac{\partial}{\partial q_j}\left(\{\tilde{p}_k,H_0 \} 
+\frac{\partial \tilde{p}_k}{\partial t}
\right),
\\
\label{wynik2}
[p_j,p_k]_{(\tilde{p}_m, \dot{q}_m)} &= 0,
\\
\label{wynik3}
[q_j,p_k]_{(\tilde{p}_m, \dot{q}_m)} &=
\frac{\partial}{\partial p_k}\left(\{\tilde{p}_j,H_0 \}
+\frac{\partial \tilde{p}_j}{\partial t}
\right),
\end{align}
\end{subequations}
where $\dot{q}_m(\bosy{q}, \bosy{p}, t)=\frac{\partial H_0}{\partial p_m}$ and we use the Lagrange bracket defined by
\begin{equation}
[x_i, x_j]_{(\psi_k, \phi_k)}=
\frac{\partial \psi_m}{\partial x_i}\frac{\partial \phi_m}{\partial x_j} -
\frac{\partial \psi_m}{\partial x_j}\frac{\partial \phi_m}{\partial x_i}
\end{equation}
with summation over $m$.
Let us remark that for one degree of freedom (one $q$ and one $p$) equations (\ref{wynik1})-(\ref{wynik3}) reduce to  one relation
\begin{equation}
\label{warunekjednowym}
\left\{\frac{\partial \tilde{p}}{\partial p},H_0 \right\}
+\frac{\partial^2 \tilde{p}}{\partial t \partial p}
=0.
\end{equation}
Equations (\ref{wynik1})-(\ref{wynik3}) do not contain $H$ and are linear with respect to $\bosy{\tilde{p}}$, so they are remarkably simpler then (\ref{dosycwazne}) and (\ref{dosycwaznedrugie}).

Once (\ref{wynik1})-(\ref{wynik3}) are solved with respect to $\bosy{\tilde{p}}$ we can use the solution to find $\left.H\right|_{\begin{subarray}{1}\bosy{q}=
\bosy{q}\\\bosy{\pi}=\bosy{\tilde{p}}(\bosy{q},\bosy{p},t)\end{subarray}}$. Using Poincar\'e lemma to the form defined by right sides of (\ref{wazne1}) and (\ref{wazne2}) we can write
\begin{equation}
\label{nowyhamiltonian}
\begin{split}
\left.H\right|_{\begin{subarray}{1}\bosy{q}=
\bosy{q}\\\bosy{\pi}=\bosy{\tilde{p}}(\bosy{q},\bosy{p},t)\end{subarray}}=&
\int_{0}^{1} 
\left(
-\{\tilde{p}_j,H_0 \}
-\frac{\partial \tilde{p}_j}{\partial t}
+\frac{\partial H_0}{\partial p_i} \frac{\partial \tilde{p}_i}{\partial q_j}
\right)
\big|_{\begin{subarray}{1}\bosy{q}= \bosy{q_0} +
\tau(\bosy{q}-\bosy{q_0})\\\bosy{p}=\bosy{p_0}+\tau(\bosy{p}-\bosy{p_0})\end{subarray}} d\tau \ (q_j -q_{0j})\\
&+
\int_{0}^{1}
\left(
\frac{\partial H_0}{\partial p_i} \frac{\partial \tilde{p}_i}{\partial p_j}
\right)
\big|_{\begin{subarray}{1}\bosy{q}= \bosy{q_0} +
\tau(\bosy{q}-\bosy{q_0})\\\bosy{p}=\bosy{p_0}+\tau(\bosy{p}-\bosy{p_0})\end{subarray}}
 d\tau \ (p_j - p_{0j})+ F(t),
\end{split}
\end{equation}
where $(\bosy{q_0},\bosy{p_0})$ is some arbitrary point for which assumptions of Poincar\'e lemma hold (the form we integrate must be well defined in a star-shaped neighborhood of $(\bosy{q_0},\bosy{p_0})$ for every moment of time $t$).

To summarize this section let us write down a general scheme for finding the new Hamiltonian equivalent to $H_0$, expressed by the same configurational coordinates $\bosy{q}$ and some new momenta $\bosy{\pi}$.
\begin{itemize}
\item First one must solve the equations (\ref{wynik1})-(\ref{wynik3}) to find $\bosy{\tilde{p}}$ i.e. possible transformation of momenta.
\item Using formula ($\ref{nowyhamiltonian}$) one should find $\left.H\right|_{\begin{subarray}{1}\bosy{q}=
\bosy{q}\\\bosy{\pi}=\bosy{\tilde{p}}(\bosy{q},\bosy{p},t)\end{subarray}}$.
\item Finally, one must find the inverse function $\bosy{\tilde{p}^{-1}}$ and express the new Hamiltonian $H$ as a function of $(\bosy{q},\bosy{\pi},t)$.
\end{itemize}

The equations presented in this section are analogous to these in \cite{santilli2}, where canonoid transformations with dependence in $\bosy{q}$ (but not in $t$) are studied.
\section{Time independent Hamiltonian}
\label{ustepohamniezodczas}
We are going to find additional conditions under which the new Hamiltonian $H$ appears to be time independent. First note that from equality
\begin{equation}
\label{defodw}
\tilde{p_j}^{-1}(\bosy{q},\bosy{\tilde{p}}(\bosy{q},\bosy{p},t),t)=p_j.
\end{equation}
one can obtain two useful relations. The first one is the well known formula for the derivatives of an inverse function
\begin{equation}
\label{pomocpoch1}
\left.\frac{\partial \tilde{p_j}^{-1}}{\partial \pi_m}\right|_{\begin{subarray}{1}\bosy{q}=
\bosy{q}\\\bosy{\pi}=\bosy{\tilde{p}}(\bosy{q},\bosy{p},t)\end{subarray}}
=
\left(\frac{\partial \bosy{\tilde{p}}}{\partial \bosy{p}}\right)^{-1}_{jm},
\end{equation}
where $\left(\frac{\partial \bosy{\tilde{p}}}{\partial \bosy{p}}\right)^{-1}$ is the inverse matrix of the invertible matrix with elements $a_{ij}=\frac{\partial \tilde{p_i}}{\partial p_j}$. The second relation is of the form
\begin{equation}
\label{pomocpoch2}
\left.\frac{\partial \tilde{p_j}^{-1}}{\partial t}\right|_{\begin{subarray}{1}\bosy{q}=
\bosy{q}\\\bosy{\pi}=\bosy{\tilde{p}}(\bosy{q},\bosy{p},t)\end{subarray}}
=
-\left(\frac{\partial \bosy{\tilde{p}}}{\partial \bosy{p}}\right)^{-1}_{jm}
\frac{\partial \tilde{p_m}}{\partial t}
\end{equation}
and can be derived by differentiation of (\ref{defodw}) with respect to $t$. Consider now the equation (\ref{dosycwazne}). If we compound it with $\bosy{\tilde{p}^{-1}}$ we obtain $\frac{\partial H}{\partial \pi_i}$ to be
\begin{equation*}
\frac{\partial H}{\partial \pi_i}=
\left.\frac{\partial H_0}{\partial p_i}\right|_{\begin{subarray}{1}\bosy{q}=
\bosy{q}\\\bosy{p}=\bosy{\tilde{p}^{-1}}(\bosy{q},\bosy{\pi},t)\end{subarray}}.
\end{equation*}
But we want our new Hamiltonian to be time independent, so $\frac{\partial^2 H}{\partial t\partial \pi_i}=0$, and consequently
\begin{equation*}
\left.\frac{\partial^2 H_0}{\partial p_i \partial p_j}\right|_{\begin{subarray}{1}\bosy{q}=
\bosy{q}\\\bosy{p}=\bosy{\tilde{p}^{-1}}(\bosy{q},\bosy{\pi},t)\end{subarray}}
\frac{\partial \tilde{p_j}^{-1}}{\partial t}+
\left.\frac{\partial^2 H_0}{\partial p_i \partial t}\right|_{\begin{subarray}{1}\bosy{q}=
\bosy{q}\\\bosy{p}=\bosy{\tilde{p}^{-1}}(\bosy{q},\bosy{\pi},t)\end{subarray}}
=0.
\end{equation*}
Then we compound above expression with $\bosy{\tilde{p}}$
\begin{equation*}
\frac{\partial^2 H_0}{\partial p_i \partial p_j}
\left.\frac{\partial \tilde{p_j}^{-1}}{\partial t}\right|_{\begin{subarray}{1}\bosy{q}=
\bosy{q}\\\bosy{\pi}=\bosy{\tilde{p}}(\bosy{q},\bosy{p},t)\end{subarray}}
+
\frac{\partial^2 H_0}{\partial p_i \partial t}
=0.
\end{equation*}
After using (\ref{pomocpoch2}) we get
\begin{equation*}
-\frac{\partial^2 H_0}{\partial p_i \partial p_j}
\left(\frac{\partial \bosy{\tilde{p}}}{\partial \bosy{p}}\right)^{-1}_{jm}
\frac{\partial \tilde{p_m}}{\partial t}
+
\frac{\partial^2 H_0}{\partial p_i \partial t}
=0.
\end{equation*}
Let $K$ be the inverse matrix of the matrix $(b_{ij})=\left(\frac{\partial^2 H_0}{\partial p_i \partial p_j}\right)$. ($K$ exists unless there are constraints imposed on velocities $\bosy{\dot{q}}$). Then we can rewrite derived formula in the final form
\begin{equation}
\label{peddostacjham}
\frac{\partial \tilde{p_l}}{\partial t}
=
\frac{\partial \tilde{p_l}}{\partial p_s}K_{si}
\frac{\partial^2 H_0}{\partial p_i \partial t}.
\end{equation}
This is a linear equation for $\bosy{\tilde{p}}$ that must be fulfilled together with (\ref{wynik1}) - (\ref{wynik3}) to lead to a time independent Hamiltonian.

Analyze equation (\ref{dosycwaznedrugie}) and perform same procedure as before. First we come over to $\bosy{\pi}$, then we differentiate the result with respect to $t$ and finally we come back to original momenta $\bosy{p}$. The result reads
\begin{multline*}
-\left\{ \frac{\partial \tilde{p_j}}{\partial t} , H_0\right\}
-\left\{\tilde{p_j}, \frac{\partial H_0}{\partial t} \right\}
-\left\{ \frac{\partial \tilde{p_j}}{\partial p_m} , H_0\right\}
\left.\frac{\partial \tilde{p}_m^{-1}}{\partial t}\right|_{\begin{subarray}{1}\bosy{q}=
\bosy{q}\\\bosy{\pi}=\bosy{\tilde{p}}(\bosy{q},\bosy{p},t)\end{subarray}}+\\
-\left\{\tilde{p_j}, \frac{\partial H_0}{\partial p_m} \right\}
\left.\frac{\partial \tilde{p}_m^{-1}}{\partial t}\right|_{\begin{subarray}{1}\bosy{q}=
\bosy{q}\\\bosy{\pi}=\bosy{\tilde{p}}(\bosy{q},\bosy{p},t)\end{subarray}}
-\frac{\partial^2 \tilde{p}_j}{\partial t^2}
-\frac{\partial^2 \tilde{p}_j}{\partial p_m \partial t}
\left.\frac{\partial \tilde{p}_m^{-1}}{\partial t}\right|_{\begin{subarray}{1}\bosy{q}=
\bosy{q}\\\bosy{\pi}=\bosy{\tilde{p}}(\bosy{q},\bosy{p},t)\end{subarray}}
=0.
\end{multline*}
Substituting (\ref{peddostacjham}) into (\ref{pomocpoch2}) we get
\begin{equation*}
\left.\frac{\partial \tilde{p}_m^{-1}}{\partial t}\right|_{\begin{subarray}{1}\bosy{q}=
\bosy{q}\\\bosy{\pi}=\bosy{\tilde{p}}(\bosy{q},\bosy{p},t)\end{subarray}}
=
-K_{mv}
\frac{\partial^2 H_0 }{\partial p_v \partial t}.
\end{equation*}
This formula and (\ref{pomocpoch2}) lead us to the following expression
\begin{multline*}
-\underbrace{\left\{ \frac{\partial \tilde{p_j}}{\partial p_m}K_{mv} \frac{\partial^2 H_0}{\partial p_v \partial t} , H_0\right\}}_{1}
-\underbrace{\left\{\tilde{p_j}, \frac{\partial H_0}{\partial t} \right\}}_{2}
+\underbrace{\left\{ \frac{\partial \tilde{p_j}}{\partial p_m} , H_0\right\}+
K_{mv}\frac{\partial^2 H_0 }{\partial p_v \partial t}}_{1}+\\
+\underbrace{\left\{\tilde{p_j}, \frac{\partial H_0}{\partial p_m} \right\}
K_{mv}\frac{\partial^2 H_0 }{\partial p_v \partial t}}_{2}
-\underbrace{\frac{\partial}{\partial t}
\left( \frac{\partial \tilde{p_j}}{\partial p_m}K_{mv} \frac{\partial^2 H_0}{\partial p_v \partial t} \right)}_{3}
+\underbrace{\frac{\partial^2 \tilde{p}_j}{\partial p_m \partial t}
K_{mv}\frac{\partial^2 H_0 }{\partial p_v \partial t}}_{3}
=0.
\end{multline*}
Now, after cancellations in groups indicated by numbered braces, we may write
\begin{multline*}
-\underbrace{\frac{\partial \tilde{p_j}}{\partial p_m} \left\{K_{mv} \frac{\partial^2 H_0}{\partial p_v \partial t} , H_0\right\}}_{1}
+\underbrace{\frac{\partial \tilde{p_j}}{\partial p_m}  \frac{\partial^2 H_0}{\partial q_m \partial t}
-\frac{\partial \tilde{p_j}}{\partial p_m} \frac{\partial^2 H_0}{\partial q_m \partial p_s}
K_{sv}\frac{\partial^2 H_0 }{\partial p_v \partial t}}_{2}+\\
-\underbrace{\frac{\partial \tilde{p_j}}{\partial p_m} \frac{\partial}{\partial t}
\left(K_{mv} \frac{\partial^2 H_0}{\partial p_v \partial t} \right)}_{3}
=0.
\end{multline*}
Finally, as $\frac{\partial \tilde{p_j}}{\partial p_m}$ are elements of invertible matrix, we get the main result
\begin{equation}
\label{warunekbezczasowosci}
\frac{\partial^2 H_0}{\partial q_i \partial t}
-\frac{\partial^2 H_0}{\partial q_i \partial p_s}
K_{sv}\frac{\partial^2 H_0 }{\partial p_v \partial t}
-\left\{K_{iv} \frac{\partial^2 H_0}{\partial p_v \partial t} , H_0\right\}
- \frac{\partial}{\partial t}
\left(K_{iv} \frac{\partial^2 H_0}{\partial p_v \partial t} \right)
=0.
\end{equation}
In this equation there is no $H$, $\bosy{\tilde{p}}$ or $\bosy{\pi}$. Consequently we interpret it as a condition that $H_0$ should satisfy to have a time independent counterpart.
\section{Examples}
\subsection{$\tilde{p_i}=p_i+\frac{\partial f}{\partial q_i}$}
One can easily check that $\tilde{p_i}=p_i+\frac{\partial f}{\partial q_i}$ is a solution to (\ref{wynik1}) -- (\ref{wynik3}) for arbitrary $f(\bosy{q},t)$ and every $H_0$. Integrating (\ref{nowyhamiltonian}) around $(\bosy{q},\bosy{p},t)$ we get
\begin{equation*}
\left.H\right|_{\begin{subarray}{1}\bosy{q}=
\bosy{q}\\\bosy{\pi}=\bosy{\tilde{p}}(\bosy{q},\bosy{p},t)\end{subarray}}=
H_0(\bosy{q},\bosy{p},t)-\frac{\partial f}{\partial t}-
H_0(\bosy{0},\bosy{0},t) + \left. \frac{\partial f}{\partial t} \right|_{\bosy{q}=0} + F(t).
\end{equation*}
Then expressing new Hamiltonian as a function of $\bosy{\pi}$ and putting $F(t)=H_0(\bosy{0},\bosy{0},t) - \left. \frac{\partial f}{\partial t} \right|_{\bosy{q}=0}$ we obtain
\begin{equation*}
H(\bosy{q},\bosy{\pi},t)=H_0(\bosy{q},\bosy{\pi} - \frac{\partial f}{\partial \bosy{q}},t)-\frac{\partial f}{\partial t}.
\end{equation*}
One quickly shows that such transformation of Hamiltonian corresponds to the standard ambiguity in Lagrangian for a system (namely adding term $\frac{\partial f}{\partial q_i} \dot{q}_i + \frac{\partial f}{\partial t}$ to Lagrangian). Notice that this transformation is canonical because it holds for every Hamiltonian.

\subsection{The damped oscillator}
The damped oscillator is a commonly used system for testing theoretical ideas (see e.g. \cite{plebanski} -- covering topics closely related to ours, or \cite{dito} and references there). Let us find a time independent Hamiltonian for it. We additionally want our new Hamiltonian to become the standard Hamiltonian of the harmonic oscillator $\left( \frac{p^2}{2}+\frac{\omega^2 q^2}{2} \right)$ if there is no damping. There exists an example, given by Havas \cite{havas}, of the time independent Hamiltonian for damped oscillator. But the Havas Hamiltonian yields quite strange form in the case when $\beta=0$. As a starting point we  use the Hamiltonian
\begin{equation}
\label{hamstd}
H_s=\frac{1}{2}\left[ e^{-2 \beta t}p_s^2 + (\beta^2 + \omega^2) e^{2 \beta t} q^2 \right],
\end{equation}
which is time dependent but for $\beta=0$ is exactly the Hamiltonian of the harmonic oscillator. After simple calculation we observe that (\ref{hamstd}) satisfies condition (\ref{warunekbezczasowosci}). In our case the equation (\ref{peddostacjham}) becomes
\begin{equation*}
\frac{\partial \tilde{p}}{\partial t}+2 \beta p_s \frac{\partial \tilde{p}}{\partial p_s}=0.
\end{equation*}
General solution to this equation is 
\begin{equation}
\label{prefabrykat1}
\tilde{p}=v\left( q, z \right),
\end{equation}
where $z=p_s e^{-2\beta t}$ and $v(q,z)$ is an arbitrary (sufficiently smooth) function of two variables. But $\tilde{p}$ must also be a solution to (\ref{warunekjednowym}). If we put (\ref{prefabrykat1}) to (\ref{warunekjednowym}) we obtain after some transformations the following relation
\begin{equation}
\label{oscyltlumtrudrow}
\frac{\partial g}{\partial q}z
-\frac{\partial g}{\partial z}
\left((\beta^2 + \omega^2)q +2\beta z \right) = 2 \beta g,
\end{equation}
where  $g(q,z)=\frac{\partial v}{\partial z}$ .
The general solution to this equation is given by
\begin{multline}
\label{rozwoscyltlumtrudrow}
g(q,z)=
\frac{1}{q^2 \omega^2 +(z+\beta q)^2} \times \\
\times
\psi\left( \frac{1}{\sqrt{q^2 \omega^2 +(z+\beta q)^2}} 
\exp \left( \frac{\beta}{\omega} \arc \tan \left( \frac{z + \beta q}{\omega q}\right) \right)
\right),
\end{multline}
where $\psi$ is an arbitrary function.
To prove this one observes that $t_1=g(q^2 \omega^2 + (z+\beta q)^2)$ and $t_2=\frac{1}{\sqrt{q^2 \omega^2 +(z+\beta q)^2}} 
\exp \left( \frac{\beta}{\omega} \arc \tan \left( \frac{z + \beta q}{\omega q}\right) \right)$ are two independent integrals of the system of equations
\begin{equation*}
\frac{dq}{z} = - \frac{dz}{(\beta^2+\omega^2)q +2\beta z} = \frac{dg}{2 \beta g}.
\end{equation*}
From (\ref{rozwoscyltlumtrudrow}) and (\ref{prefabrykat1}) we have
\begin{multline}
\label{bardzoogolnerozwot}
\frac{\partial \tilde{p}}{\partial p_s}=
\frac{e^{-2\beta t}}{q^2 \omega^2 +(p_s e^{-2\beta t}+\beta q)^2} \times \\
\times
\psi\left( \frac{1}{\sqrt{q^2 \omega^2 +(p_s e^{-2\beta t}+\beta q)^2}} 
\exp \left( \frac{\beta}{\omega} \arc \tan \left( \frac{p_s e^{-2\beta t} + \beta q}{\omega q}\right) \right)
\right).
\end{multline}
But for $\beta=0$ the Hamiltonian should remain unchanged so without damping one should have $\tilde{p}(q,p_s,t)=p_s$ and  $\frac{\partial \tilde{p}}{\partial p_s}=1$. Let us write (\ref{bardzoogolnerozwot}) for  $\beta=0$:
\begin{equation*}
\frac{\partial \tilde{p}}{\partial p_s}=
\frac{1}{q^2 \omega^2 +p_s^2} 
\psi\left( \frac{1}{\sqrt{q^2 \omega^2 +p_s^2}} 
\right).
\end{equation*}
This allows us to choose $\psi(\xi)=\xi^{-2}$. It is not the only possible choice as  $\psi$ can also be a function of $\beta$ and above equation gives us information only for  $\beta=0$. Using this form of $\psi$ we obtain
\begin{equation}
\label{otnaszatransfom}
\frac{\partial \tilde{p}}{\partial p_s}=
e^{-2\beta t} 
\exp \left( - \frac{2 \beta}{\omega} \arc \tan \left( \frac{p_s e^{-2\beta t} + \beta q}{\omega q}\right) \right).
\end{equation}
In what follows we use the following notation:
\begin{itemize}
\item $B(a,b)=\int_0^1 t^{a-1}(1-t)^{b-1} dt$ is the Euler beta function,
\item $B(z,a,b)=\int_0^z t^{a-1}(1-t)^{b-1} dt$ is the incomplete beta function,
\item $I(z,a,b)=B(z,a,b)/B(a,b)$ is the regularized incomplete beta function,
\item $I^{-1}(z,a,b)$ is the inverse of the regularized incomplete beta function, i.e. function providing solution of equation $z=I(s,a,b)$, with respect to $s$,
\item $\tau^+=1+\frac{i \beta}{\omega},$
\item $\tau^-=1-\frac{i \beta}{\omega}.$
\end{itemize}
Using the relation $\arc \tan (x) = \frac{1}{2} i \log \left( \frac{i+x}{i-x} \right)$ and integrating we get
\begin{multline*}
\tilde{p}(q,p_s,t)=2 i \omega q B(\tau^-,\tau^+) \times\\
\times
\left[ 
I\left( \frac{1}{2}\left( \tau^- + \frac{p_s e^{-2 \beta t}}{i \omega q}\right), \tau^-,\tau^+ \right) - 
I\left( \frac{1}{2} \tau^- , \tau^-,\tau^+ \right)
\right].
\end{multline*}
In this result we chose integration constant taking $\tilde{p}(q,0,t)=0$. Note that $\tilde{p}$ is real. The inverse transformation can be written as follows (we abandon $\pi$ as a symbol for new momentum because it could be confusing in context of trigonometric functions; instead we use $p_n$).
\begin{multline*}
\tilde{p}^{\ -1} (q,p_n,t)=e^{2 \beta t} \times \\
\times
\left( 
2 i \omega q \left[
I^{-1} \left( \frac{p_n}{2 i \omega q B(\tau^-,\tau^+)} + I\left( \frac{1}{2} \tau^- , \tau^-,\tau^+ \right),\tau^-,\tau^+ \right)
-\frac{1}{2}
\right]
-\beta q
\right) 
\end{multline*}
Applying the  formula ($\ref{nowyhamiltonian}$) one finds
\begin{equation*}
\left.H\right|_{\begin{subarray}{1}q=
q\\
\pi=\tilde{p}(q,p_s,t)\end{subarray}}=
\frac{1}{2} e^{-2 \beta t} \frac{\partial \tilde{p}}{\partial p_s}
\left( (p_s+e^{2 \beta t} \beta q)^2 + e^{4 \beta t} q^2 \omega^2 \right) 
\end{equation*}
The form we have integrated is not well defined in $(0,0,t)$ but one can check that the above result holds true. We left term $\frac{\partial \tilde{p}}{\partial p_s}$ to shorten further calculations. Let us pass to momentum $p_n$
\begin{multline*}
H=
\frac{1}{2} e^{-2 \beta t} \left. \frac{\partial \tilde{p}}{\partial p_s} \right|_{p_s = \tilde{p}^{\ -1}(q,p_n,t)} q^2 \omega^2 \times \\
\times
\left( 
1-\left[
2
I^{-1} \left( \frac{p_n}{2 i \omega q B(\tau^-,\tau^+)} + I\left( \frac{1}{2} \tau^- , \tau^-,\tau^+ \right),\tau^-,\tau^+ \right)
-1
\right]^2
\right).
\end{multline*}
But $\left. \frac{\partial \tilde{p}}{\partial p_s} \right|_{p_s = \tilde{p}^{\ -1}(q,p_n,t)}=1/\frac{\partial \tilde{p}^{\ -1}}{\partial p_n}$. By using the formula for derivative of $I^{-1}$ (see~\cite{wzormath})
\begin{equation}
\label{pochinvbetaregul}
\frac{\partial I^{-1} (z,a,b)}{\partial z} = B(a,b)(1-I^{-1}(z,a,b))^{1-b} (I^{-1}(z,a,b))^{1-a}
\end{equation}
we obtain
\begin{equation*}
\frac{\partial \tilde{p}^{\ -1}}{\partial p_n}=e^{2 \beta t}
\left(
\frac{1}
{I^{-1} \left( \frac{p_n}{2 i \omega q B(\tau^-,\tau^+)} + I\left( \frac{1}{2} \tau^- , \tau^-,\tau^+ \right),\tau^-,\tau^+ \right)}
-1
\right)^{1-\tau^+}.
\end{equation*}
Finally
\begin{multline}
\label{otnowyhamiltonnc}
H(q,p_n)= \\
=2 q^2 \omega^2 
\left[ 
1- I^{-1} \left( \frac{p_n}{2 i \omega q B(\tau^-,\tau^+)} + I\left( \frac{1}{2} \tau^- , \tau^-,\tau^+ \right),\tau^-,\tau^+ \right)
\right]^{\tau^+} \times \\
\left[ 
I^{-1} \left( \frac{p_n}{2 i \omega q B(\tau^-,\tau^+)} + I\left( \frac{1}{2} \tau^- , \tau^-,\tau^+ \right),\tau^-,\tau^+ \right)
\right]^{\tau^-}.
\end{multline}
One can check by direct calculations that our new Hamiltonian is a real function of $q$ and $p_n$ and it gives equation of motion of the damped oscillator. For $\beta=0$ (and consequently $\tau^+=\tau^-=1$) (\ref{otnowyhamiltonnc}) reduces to $\frac{1}{2}\left(p_n^2 + q^2 \omega^2 \right)$. Notice that there is a discontinuity in (\ref{otnowyhamiltonnc}) for $q=0$.

\section{Conclusions}
We have developed method for generating new Hamiltonians by changing canonical momenta. We have applied this method to find the condition (\ref{warunekbezczasowosci}) for existence of a time independent Hamiltonian and to show how to construct it. The ambiguity in Hamiltonians is especially intriguing in quantum context. We have given example of the damped oscillator and have found the new Hamiltonian for it. We believe that this new Hamiltonian can yield some interesting results at quantum level.

\section{Acknowledgments}
The author would like to thank Professor Maciej Przanowski for his remarks and support.

\end{document}